# MD-Recon-Net: A Parallel Dual-Domain Convolutional Neural Network for Compressed Sensing MRI

Maosong Ran, Wenjun Xia, Yongqiang Huang, Zexin Lu, Peng Bao, Yan Liu, Huaiqiang Sun, Jiliu Zhou, *Senior Member, IEEE*, and Yi Zhang*, *Senior Member, IEEE*

***Abstract*—Compressed sensing magnetic resonance imaging (CS-MRI) is a theoretical framework that can accurately reconstruct images from undersampled *k*-space data with a much lower sampling rate than the one set by the classical Nyquist-Shannon sampling theorem. Therefore, CS-MRI can efficiently accelerate acquisition time and relieve the psychological burden on patients while maintaining high imaging quality. The problems with traditional CS-MRI reconstruction are solved by iterative numerical solvers, which usually suffer from expensive computational cost and the lack of accurate handcrafted priori. In this paper, inspired by deep learning's (DL's) fast inference and excellent end-to-end performance, we propose a novel cascaded convolutional neural network called MD-Recon-Net to facilitate fast and accurate MRI reconstruction. Especially, different from existing DL-based methods, which operate on single domain data or both domains in a certain order, our proposed MD-Recon-Net contains two parallel and interactive branches that simultaneously perform on *k*-space and spatial-domain data, exploring the latent relationship between *k*-space and the spatial domain. The simulated experimental results show that the proposed method not only achieves competitive visual effects to several state-of-the-art methods, but also outperforms other DL-based methods in terms of model scale and computational cost.**

## I. Introduction

Magnetic resonance imaging (MRI) is a mainstream medical imaging technology that can reveal both internal anatomical structures and physiological functions noninvasively. Although MRI is superior to other medical imaging modalities in terms of both soft-tissue contrast and resolution, MRI's sampling process in *k*-space suffers from prolonged acquisition time due to physiological and hardware constraints. Even worse, such long sampling time causes patients discomfort and potentially leads to motion artifacts. These limitations make MRI unsuitable for time-critical diagnostics such as stroke. Meanwhile, hybrid PET/MR imaging is growing in several clinical applications, such as early diagnosis for tumors and cardio-cerebral diseases [1-3]. Since the PET scan is much faster than MR, the imaging speed of PET/MR is dominated by the MR scan. As a result, efficient reconstruction algorithms that accelerate MRI are in high demand.

Over the past few years, various efforts have been made to develop advanced reconstruction algorithms to reduce the acquisition time in MRI. These methods fall into two categories: the first one is hardware-based parallel MRI (pMRI) [4], which uses phased array coils containing multiple independent receiver channels [5]. Each coil has sensitivity to measure raw data from an individual tissue type, which requires specific

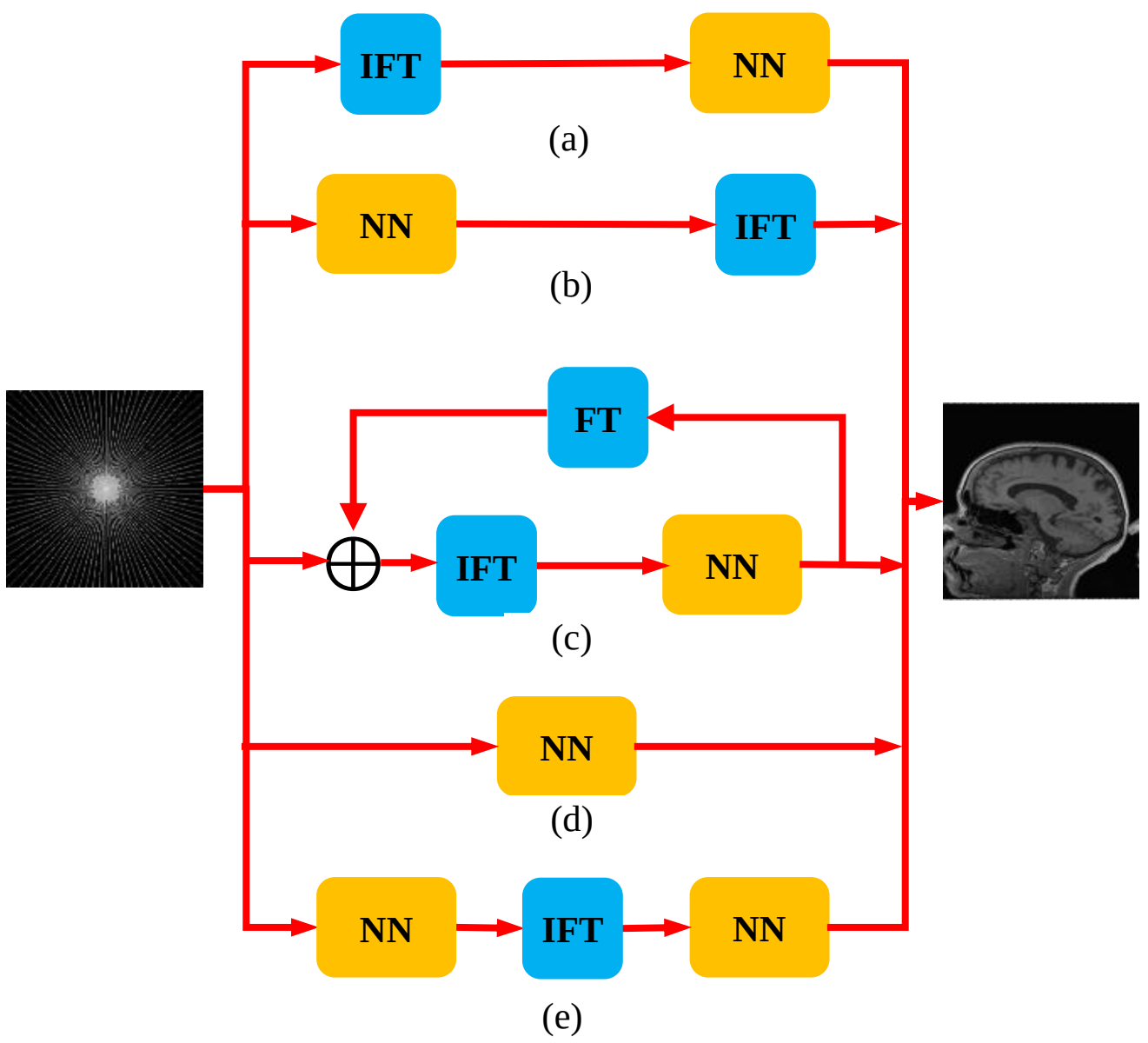

**Fig.1** DL-based methods for fast MRI. (a) Spatial-domain based methods; (b) *k*-space based methods; (c) iteration unrolling methods; (d) directly mapping from *k*-space to image; (e) cross-domain reconstruction. NN: Neural Network, IFT: Inverse Fourier Transform.

This work was supported in part by the National Natural Science Foundation of China under Grant 61671312, in part by the Sichuan Science and Technology Program under Grant 2018HH0070. (Corresponding author: Yi Zhang)

M. Ran, W. Xia, Y. Huang, Z. Lu, P. Bao, J. Zhou and Y. Zhang are with the College of Computer Science, Sichuan University, Chengdu 610065, China (e-mail: maosongran@gmail.com; xwj90620@gmail.com; hyq_tsmotlp@163.com; zexinlu.scu@gmail.com; pengbao7598@gmail.com; zhoujl@scu.edu.cn; yzhang@scu.edu.cn )

Y. Liu is with the School of Electrical Engineering Information, Sichuan University, Chengdu 610065, China(e-mail: liuyan77@scu.edu.cn)

H. Sun is with the Department of Radiology, West China Hospital of Sichuan University, Chengdu 610041, China (e-mail: sunhuaiqiang@scu.edu.cn)

pMRI reconstruction algorithms. These pMRI reconstruction methods can be classified into three groups: (a) image domain based methods, such as SENSE [6, 7] and its variants [8-11]; (b) *k*-space based methods, such as SMASH [12-14] and GRAPPA [15]; and (c) combinations of the previous two kinds of methods, such as SPACE-RIP [16] and SPIRiT [17]. Despite wide clinical application, pMRI still has modest acceleration rates. The second acceleration strategy is based on the theory of compressed sensing (CS) [18], called compressed sensing magnetic resonance imaging (CS-MRI) [8, 19]. Because the acceleration rate is proportional to the sampling ratio, CS-MRI tries to accelerate reconstruction with undersampled *k*-space data at a rate much lower than the one set by the classical Nyquist-Shannon sampling theorem. In the sense of CS-MRI, the image to reconstruct is assumed to be sparse after application of a certain sparsifying transform. Typical transforms include Fourier transform (FT), wavelet, total variation (TV), and low-rank[8, 18-20]. Recently, two data-driven based CS methods, dictionary learning and transform learning, have been introduced to improve the performance of CS-MRI [21-23]. Although these methods have achieved considerable success, several critical drawbacks still hamper the clinical practice of CS-MRI: (i) the iterative procedure is time consuming. A complex regularization term would even further aggravate the cost; (ii) parametric adjustment is exhausting for users; and (iii) the handcrafted regularization terms cannot ensure consistently superior performance for all scanning protocols and patients because of the lack of accurate prior information.

In recent years, the success of deep learning (DL) has suggested a new orientation in fast MRI reconstruction. DL-based fast MRI has gained much attention, and many new methods have emerged. These methods can be roughly divided into five groups according to data processing and specific pipeline. The basic flowcharts are illustrated in Fig. 1. The first group as shown in Fig. 1(a) is post-processing algorithms that use the inverse Fourier transform (IFT) to obtain an initial image as the first input to the network. In this group, the network model acts a role of an image-to-image mapping function [24-28]. For examples, [25, 28] are the earliest works to introduce generative adversarial network (GAN) into post-processing based fast MRI reconstruction. [28] added a data consistency layer to ensure the points on the realistic data manifold. [24] used a deeper generator and discriminator networks with cyclic data consistency loss to further enhance the imaging quality. This kind of method is the mainstream of current DL-based models, as it is convenient to insert into the current workflow of commercial scanners. In contrast, the second class as shown in Fig. 1(b) directly deals with the undersampled *k*-space data using a neural network, and after that, IFT is applied to obtain the final results [29]. Because any artifacts introduced by the network may spread to the whole reconstructed image, this kind of method has not been widely studied yet. The third branch in Fig. 1(c) comprises the iteration unrolling methods [30-34]. Specifically, in [34], the data consistency lay was embedded into the unrolling iteration network and the same group improved the former network by introducing dilated convolution and stochastic architecture [33]. In these methods, different iterative numerical solvers are treated as different recurrent networks, and the learned regularization terms constrain the image in terms of each intermediate reconstructed result. Fig. 1(d) includes methods that directly learn the image from the undersampled *k*-space data [35]. Fully connected layers are usually needed for this kind of model, and the network scale is generally huge. Until now, all the methods mentioned above have all performed in a single domain and have not fully explored the latent relationship between *k*-space and the spatial domain. The last pipeline illustrated in Fig. 1(e) has gained much attention very recently. This type of method attempts to explore the information in both *k*-space and the spatial domain [36-38]. It usually adopts two cascaded networks performing on *k*-space and spatial-domain data, with IFT employed to build the bridge between the two networks. Currently, state-of-the-art performance is achieved using such dual-domain based methods. However, existing dual-domain methods [36-38] process the *k*-space and spatial-domain data sequentially, which implicitly adds a certain priority priori into the reconstruction and may ignore the internal interplay between both domains. In this paper, to essentially address the intrinsic relation between the *k*-space and spatial domains, we propose a novel MRI Dual-domain Reconstruction Network (MD-Recon-Net) to accelerate magnetic resonance imaging. Different from current methods, the proposed MD-Recon-Net contains two parallel and interactive branches that simultaneously operate on *k*-space and spatial-domain data. Data consistency layers are included to improve performance further. At the end, dual-domain fusion layers combine the results from the two branches.

The rest of this paper is organized as follows. Related works and the proposed network are described in Section 2. The experiments and evaluation are presented in Section 3. The final section concludes this paper.

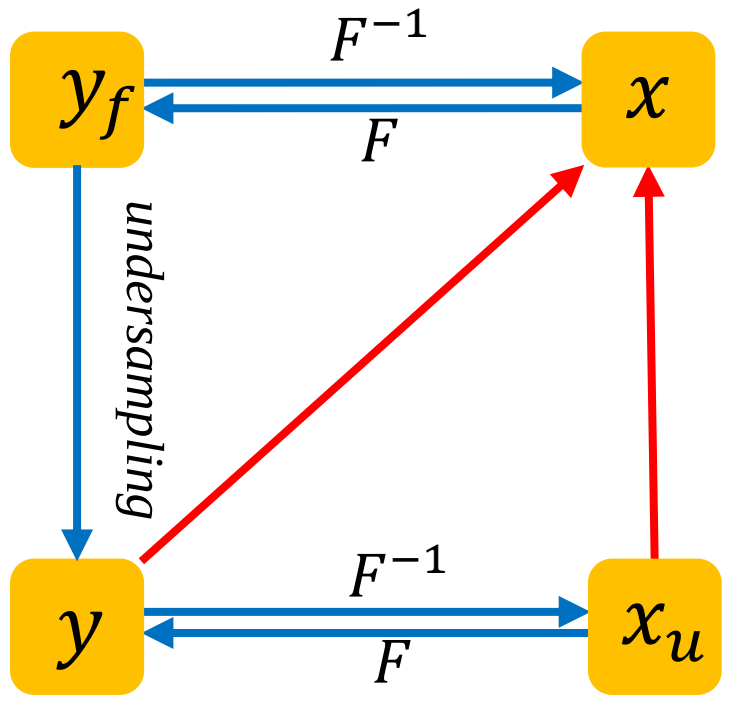

**Fig.2** Relationship among different forms of measurements. The red arrows denote the accelerated MRI algorithms. $F$ and $F^{-1}$ denote the 2D IFT and IFT respectively.

## II. METHODS

### A. *Notations*

We denote the fully sampled *k*-space data as $y_f \in \mathbb{C}^{m\times n}$ and the corresponding undersampled measurements as $y \in \mathbb{C}^{m\times n}$, where $m$ and $n$ are the image size. The zero-filling solution reconstructed from $y$ is denoted by $x_u \in \mathbb{C}^{m\times n}$, and the reconstructed image from $y_f$ is $x \in \mathbb{C}^{m\times n}$. In general, the

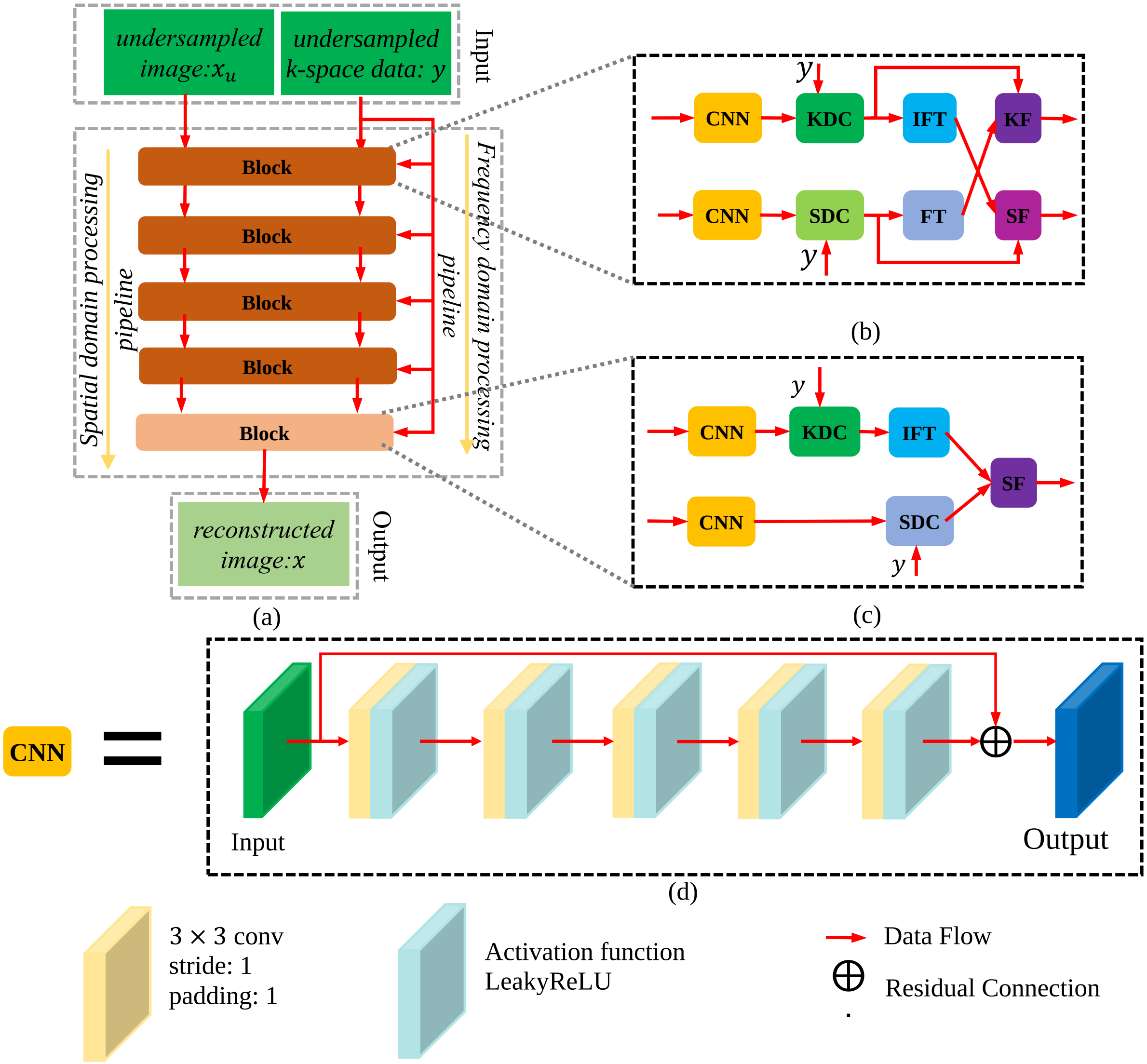

**Fig. 3** The structure of the proposed method. (a) Overall architecture of the proposed network, (b) the details of the first four processing blocks, (c) the details of the last processing blocks, (d) the details of CNN in each block. CNN: convolutional neural network, KDC: *k*-space data consistency, SDC: spatial data consistency, FT: Fourier Transform, IFT: Inverse Fourier Transform, KF: *k*-space data fusion layer, SF: spatial data fusion layer.

problem of image reconstruction can be approximately treated as solving a linear system formulated as:

$$y = U \odot y_f + \epsilon = U \odot Fx + \epsilon = F_u x + \epsilon \qquad (1)$$

where $F$ denotes the 2D FT operator, $U \in R^{m\times n}$ represents the binary undersampling mask, $F_u = U \odot F$ denotes the undersampling Fourier encoding operator, $\odot$ is element-wise multiplication, and $\epsilon$ is acquisition noise. Fig. 2 illustrates the relationships among different data forms, and the red arrows indicate the accelerated MRI algorithms on which we focus here.

### B. CS-MRI

To solve the ill-posed problem in (1), classic model-based CS-MRI [18, 19] constrains the solution space by exploiting some prior knowledge, and the associated optimization problem can be expressed as the following variational minimization:

$$\min_x \frac{1}{2}\|F_u x - y\|_2^2 + \lambda \Re(x) \qquad (2)$$

where the first term represents data fidelity, which guarantees consistency between the reconstruction result and the original undersampled *k*-space data, $\Re(x)$ denotes the regularization term that constrains the least squares data fidelity term, and $\lambda \geq 0$ is a balancing parameter that controls the tradeoff between the two terms. Especially, $\Re(x)$ is usually an $l_0$ norm or $l_1$ norm in a certain sparsifying transform field, and some typical regularizers include FT, wavelet, total variation (TV), and low-rank[8, 18-20].

DL-based CS-MRI combines the deep learning with CS-MRI [24-27, 29, 34, 38-40], which makes the reconstructed image and the corresponding fully sampled image as close as possible by optimizing the parameter set $\theta$ of the neural network. This method can be represented as:

$$\min_x \lambda\|f_{nn}(z|\theta) - x\|^2 + \frac{1}{2}\|F_u x - y\|_2^2 \qquad (3)$$

where $f_{nn}$ is the network model with parameter set $\theta$, $z$ is the input of the model, either $y$ or $x_u$, and $f_{nn}(z|\theta)$ is the output of the model, which denotes the predicted reconstruction result. The critical part is how to design the architecture of network in (3) as in [40] or even replace the energy minimization process with the training process of a neural network, as in [24, 25, 36]. In this paper, we focus on the second strategy.

## C. Proposed Method

In this section, we elaborate on the details of the proposed MD-Recon-Net and introduce the network's key components.

### 1) Network Architecture

The detailed architecture of the proposed MD-Recon-Net is illustrated in Fig. 3(a). It consists of five basic blocks, and all blocks share the same structure except for the last block. The network takes both undersampled $k$-space data $y$ and zero-filling reconstruction $x_u$ as input and predicts the full-sampling reconstruction $x$. The proposed network contains two branches to process the $k$-space and spatial data separately.

The overall structures of the basic processing blocks are illustrated in Fig. 3(b) and (c). The block structure is key to the proposed method and is the main contribution of this paper. The first four blocks share the same structure, which is shown in Fig. 3(b). This block contains the following components: CNN, FT, IFT, $k$-space data consistency (KDC), spatial data consistency (SDC), $k$-space fusion (KF), and spatial fusion (SF) layers. Each block accepts two inputs, $k$-space and spatial data, and contains two CNN modules, which are used to extract and recover the features in both domains. The details of the CNN are shown in Fig. 3(d). It consists of 5 convolutional layers with 32, 32, 32, 32, and 2 filters, respectively. Residual connection [41] is introduced to accelerate the training procedure and preserve more detail. Because MR data are complex valued, two channels are used to represent the real and imaginary parts, respectively. The input of the first convolutional layer contains two channels, and the last convolutional layer only contains two filters. All kernels are set to $3 \times 3$ and are followed by a LeakyReLU unit with negative slope $1e^{-2}$. The stride is set to 1, and we set $padding = 1$ to keep the dimensions of the input and output consistent.

The outputs of the two CNN modules are fed into two different data consistency modules, KDC and SDC, to impose data constraints on the intermediate results in both domains. The details of KDC and SDC are given in section II.C.2. After that, to fuse the results from different domains, FT and IFT are applied to the spatial and $k$-space data, respectively. Then, the outputs of KDC and FT are fed into the KF module, and the outputs of SDC and IFT are fed into the SF module. The details of KF and SF are elaborated in section II.C.3. Finally, the KF and SF modules output the intermediate reconstructed results. The last block, shown in Fig. 3(c), is similar to the previous ones in Fig. 3(b), with the only difference occurring after the data consistency layers. Because this block exports the final reconstructed images instead of the dual-domain results, the FT and KF modules are removed, and the final result is obtained after spatial fusion.

For simplicity, mean square error (MSE) is adopted as the proposed network's loss function, and it is defined as:

$$L = \frac{1}{N}\sum_{i=1}^{N} \|x_i - \hat{x}_i\|_2^2 \qquad (4)$$

where $\hat{x}$ is the predicted MR image of the network, and $N$ denotes the total number of samples.

### 2) Data Consistency Module

A data consistency layer is used to impose the constraints from the original measurements. We follow the idea of [34] to incorporate the data fidelity into the neural network, and we adopt a formula for KDC according to the closed-form solution of (3), as follows:

$$s_{rec}(j) = \begin{cases} \hat{y}(j) & if\ j \notin \Omega \\ \dfrac{\hat{y} + \gamma y(j)}{1+\gamma} & if\ j \in \Omega \end{cases} \qquad (5)$$

where $j$ represents the index of the vectorized representation of $k$-space data, $\hat{y}$ denotes the predicted $k$-space data from the previous CNN module, $y$ denotes the undersampled $k$-space data, $\Omega$ is the sampling index set, and $\gamma$ is a hyperparameter.

In (5), if the $k$-space coefficients are not sampled ($j \notin \Omega$), the value predicted by the CNN module is used; for the sampled entries, this is a linear weighted summation between the CNN predicted and original sampled data. SDC plays a similar role to KDC but performs in spatial domain. It can be implemented easily by adding FT and IFT before and after SDC. Because of their simple expressions, the forward and backward passes of KDC and SDC can be easily derived. Please refer to [34] for more details.

### 3) Data Fusion Module

According to the previous study [38] and our experiments, it can be noticed that although spatial domain network generally has better performance than $k$-space domain network, some details can only be recovered by $k$-space domain network. Based on this observation, the proposed MD-Recon-Net tries to take advantage of the merits in dual domains and a parallel interactive architecture with two branches is used. One of the main contributions in this paper is that the data from both domains are not separate, but interactive via the fusion modules.

In Fig.3, two types of fusion modules, KF and SF, are used. The formulae of two modules are same and the difference lies in the used data forms. KF and SF modules respectively deal with the $k$-space and spatial data. KF module takes the outputs of previous KDC and FT as inputs and SF module takes the outputs of previous SDC and IFT as inputs. The computations of KF and SF are unified with a linear combination between two inputs and can be formulated as:

$$A = \frac{1}{1+\mu}A_1 + \frac{\mu}{1+\mu}A_2 \qquad (6)$$

where $A$ is the output of fusion module, $A_1$ and $A_2$ are the inputs, $\mu$ is the balancing factor.

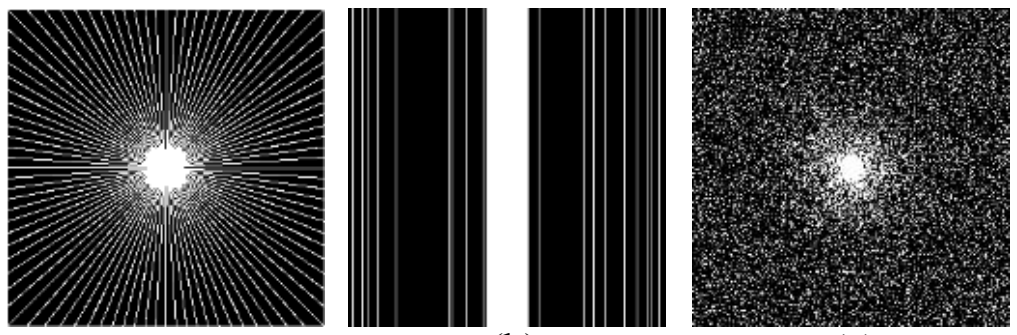

(a) (b) (c)

**Fig. 4** Examples of three types of under-sampling masks. (a) Pseudo radial sampling mask, (b) Cartesian sampling mask, (c) Gaussian random sampling mask.

## III. Experimental Design and Representative Results

### A. Experimental Simulation Configuration

The public brain MR raw data set—the *Calgary-Campinas* dataset[1], which comes from a clinical MR scanner (Discovery MR750; GE Healthcare, Waukesha, WI)—was used to train and test our proposed model. In total, 4,524 slices from 25 subjects were randomly selected to form the training set, and the testing set was composed of 1,700 slices from 10 other subjects. The size of the acquisition matrix is $256 \times 256$.

Raw MR data are complex valued, but CNNs can only handle real numbers. There are two common strategies to enable CNNs to process complex numbers [25]: (a) using two separate channels to represent the real and imaginary parts, and (b) using the magnitude of the complex number as the network input. In this paper, the first strategy was adopted.

The Adam optimizer [42] was used to optimize the network, and its parameters were set as $\alpha = 5e^{-5}, \beta_1 = 0.9, \beta_2 = 0.999$. The initial learning rate was $5e^{-5}$. The parameters $\gamma$ in (5) an d $\mu$ in (6) were trained as network parameters. The PyTorch fr amework was used for model implementation, and the training was performed on a graphics processing unit (GTX 1080 Ti).

Our codes for this work are available on https://github.com/Deep-Imaging-Group/MD-Recon-Net.

For undersampling strategies, three different types of simulated masks were tested: pseudo radial sampling, Cartesian sampling, and 2D random sampling (see Fig. 4). For pseudo radial sampling, 10%, 20%, and 25% sampling rates were tested, and only a 20% sampling rate was used for the other two types.

To evaluate the performance of the proposed MD-Recon-Net, its performance was compared with that of five state-of-the-art methods: DLMRI [21], PANO [43], ADMM-CSNet [30], DIMENSION [36] and DAGAN [25]. DLMRI[2] and PANO[3] are two typical CS-MRI methods based on dictionary learning and nonlocal means, respectively. ADMM-CSNet[4] is a recently proposed iteration unrolling method. DAGAN[5] and DIMENSION are two post-processing network models that perform on the spatial and dual domains, respectively. All codes of the compared methods were downloaded from the authors' websites or implemented strictly according to the original papers.

To quantitatively evaluate the performance of different methods, two metrics, peak signal to noise ratio (PSNR) and structural similarity index measure (SSIM), were employed, with the following definitions

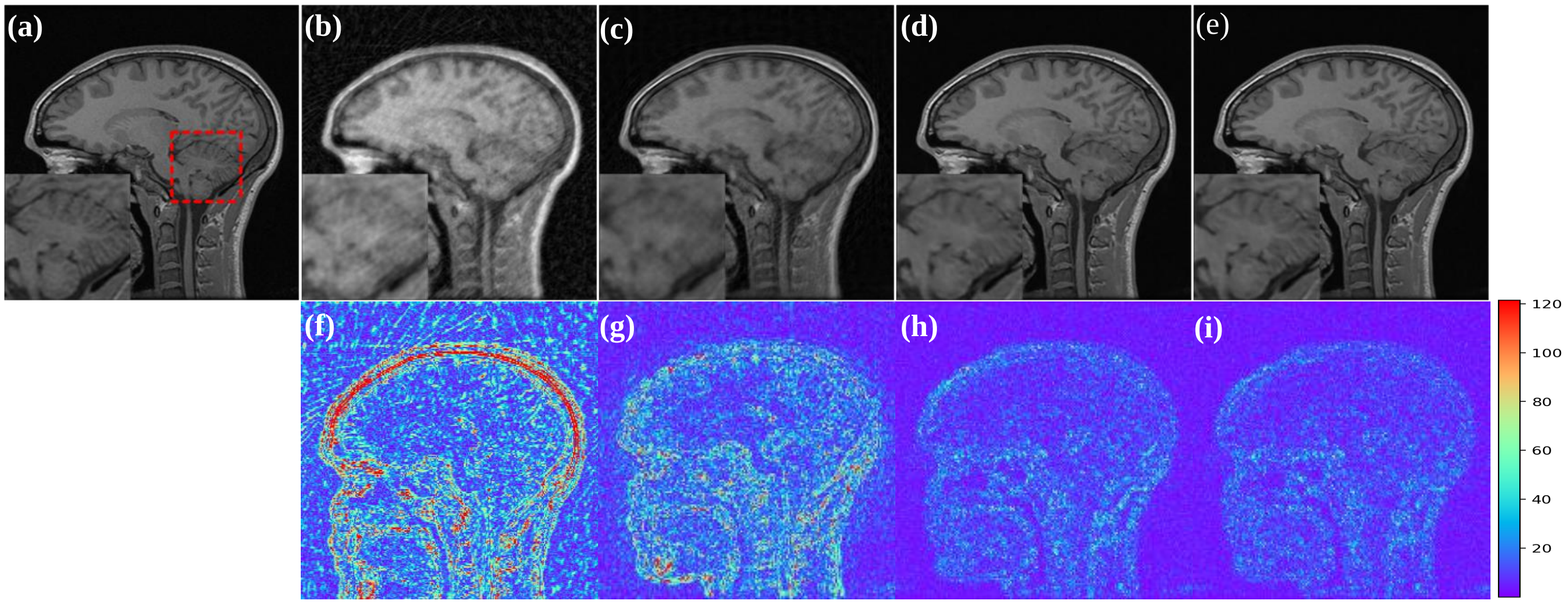

**Fig.5** Representative visual results of MD-Recon-Net and its variants. (a) Full-sampling image; (b) zero-filling image (PSNR=27.1055, SSIM=0.7298); (c) MD-Recon-Net-k (PSNR=31.0347, SSIM=0.8608); (d) MD-Recon-Net-s (PSNR=36.9282, SSIM=0.9565); (e) MD-Recon-Net(**PSNR=37.9117, SSIM=0.9625**); (f) error map of zero-filling, (g) error map of MD-Recon-Net-k; (h) error map of MD-Recon-Net-s; (i) error map of MD-Recon-Net;

**Table I**
Statistical quantitative results (PSNR and SSIM) of MD-Recon-Net and its variations for the test set

| | Cartesian | Random | Radial | | |
|---|---|---|---|---|---|
| | 20% | 20% | 10% | 20% | 25% |
| Zero-filling | 24.62±1.35 | 25.57±0.86 | 24.18 ±1.43 | 27.56±1.55 | 28.84±1.59 |
| | 0.69±0.06 | 0.67±0.04 | 0.61±0.05 | 0.75±0.04 | 0.79±0.03 |
| MD-Recon-Net-k | 26.38±0.88 | 29.33±0.82 | 27.33±0.94 | 31.66±0.97 | 33.25±0.94 |
| | 0.74±0.03 | 0.79±0.01 | 0.74±0.02 | 0.87±0.01 | 0.90±0.01 |
| MD-Recon-Net-s | 31.22±1.46 | 37.50±1.66 | 32.37±1.62 | 37.42±1.42 | 39.85±1.35 |
| | 0.88±0.02 | **0.95±0. 01** | 0.90±0.02 | 0.95±0.01 | 0.97±0.00 |
| MD-Recon-Net | **32.00±1.38** | **38.11±1.59** | **32.92±1.50** | **38.26±1.35** | **39.85±1.18** |
| | **0.90±0.02** | **0.95±0.01** | **0.91±0.02** | **0.96±0.01** | **0.97±0.01** |

[1] https://sites.google.com/view/calgary-campinas-dataset/home/mr-reconstruction-challenge
[2] http://www.ifp.illinois.edu/~yoram/DLMRI-Lab/DLMRI.html
[3] http://csrc.xmu.edu.cn/project/CS_MRI_PANO/Codes_for_PANO.rar
[4] https://github.com/yangyan92/ADMM-CSNet
[5] https://github.com/nebulaV/DAGAN

**Table II**

Statistical quantitative results (PSNR, SSIM and OQ) of different methods for the test set

| | Cartesian | Random | Radial | | | Average |
|---|---|---|---|---|---|---|
| | 20% | 20% | 10% | 20% | 25% | Execution Time(s) |
| Zero-filling | 24.62±1.35<br>0.69±0.06<br>1.03±0.03 | 25.57±0.86<br>0.67±0.04<br>1.10±0.09 | 24.18 ±1.43<br>0.61±0.05<br>1.07±0.06 | 27.56±1.55<br>0.75±0.04<br>1.13±0.12 | 28.84±1.59<br>0.79±0.03<br>1.17±0.14 | |
| PANO | 28.71±1.62<br>0.81±0.04<br>2.77±0.38 | 34.78±2.13<br>0.92±0.02<br>3.03±0.37 | 30.46±2.11<br>0.85±0.04<br>2.96±0.53 | 35.67±1.75<br>0.94±0.01<br>3.24±0.67 | 37.69±1.63<br>0.96±0.01<br>3.56±0.58 | 43.7738 |
| DLMRI | 27.26±1.68<br>0.78±0.05<br>2.68±0.46 | 30.98±1.65<br>0.85±0.03<br>2.97±0.67 | 28.29±1.90<br>0.79±0.05<br>2.81±0.78 | 32.16±1.59<br>0.88±0.02<br>3.01±0.85 | 33.57±1.51<br>0.91±0.01<br>3.18±0.45 | 253.3191 |
| ADMM-CSNet | 30.25±1.61<br>0.86±0.03<br>3.01±0.77 | 35.74±1.91<br>0.93±0.01<br>3.34±0.46 | 31.32±1.90<br>0.88±0.03<br>3.10±0.78 | 35.44±1.71<br>0.94±0.01<br>3.37±0.53 | 39.28±1.49<br>0.97±0.00<br>3.94±1.03 | 1.2643 |
| DAGAN | 28.89±1.33<br>0.83±0.03<br>2.77±0.78 | 28.89±1.77<br>0.80±0.04<br>2.90±0.81 | 27.97±1.62<br>0.78±0.03<br>2.79±0.88 | 32.08±1.89<br>0.89±0.02<br>3.07±0.97 | 33.34±1.88<br>0.91±0.02<br>3.21±0.86 | 0.0400 |
| DIMENSION | 31.38±1.74<br>0.89±0.03<br>3.34±0.67 | 35.86±2.30<br>**0.95±0.01**<br>3.47±0.89 | 32.13±1.86<br>0.90±0.02<br>3.40±1.01 | 36.21±2.26<br>0.96±0.01<br>3.78±0.78 | 37.59±2.68<br>**0.97±0.01**<br>3.97±0.97 | 0.1835 |
| MD-Recon-Net | **32.00±1.38**<br>**0.90±0.02**<br>**3.56±0.77** | **38.11±1.59**<br>**0.95±0.01**<br>**3.83±0.45** | **32.92±1.50**<br>**0.91±0.02**<br>**3.56±0.89** | **38.26±1.35**<br>**0.96±0.01**<br>**3.93±0.67** | **39.85±1.18**<br>**0.97±0.01**<br>**3.99±0.78** | **0.0116** |

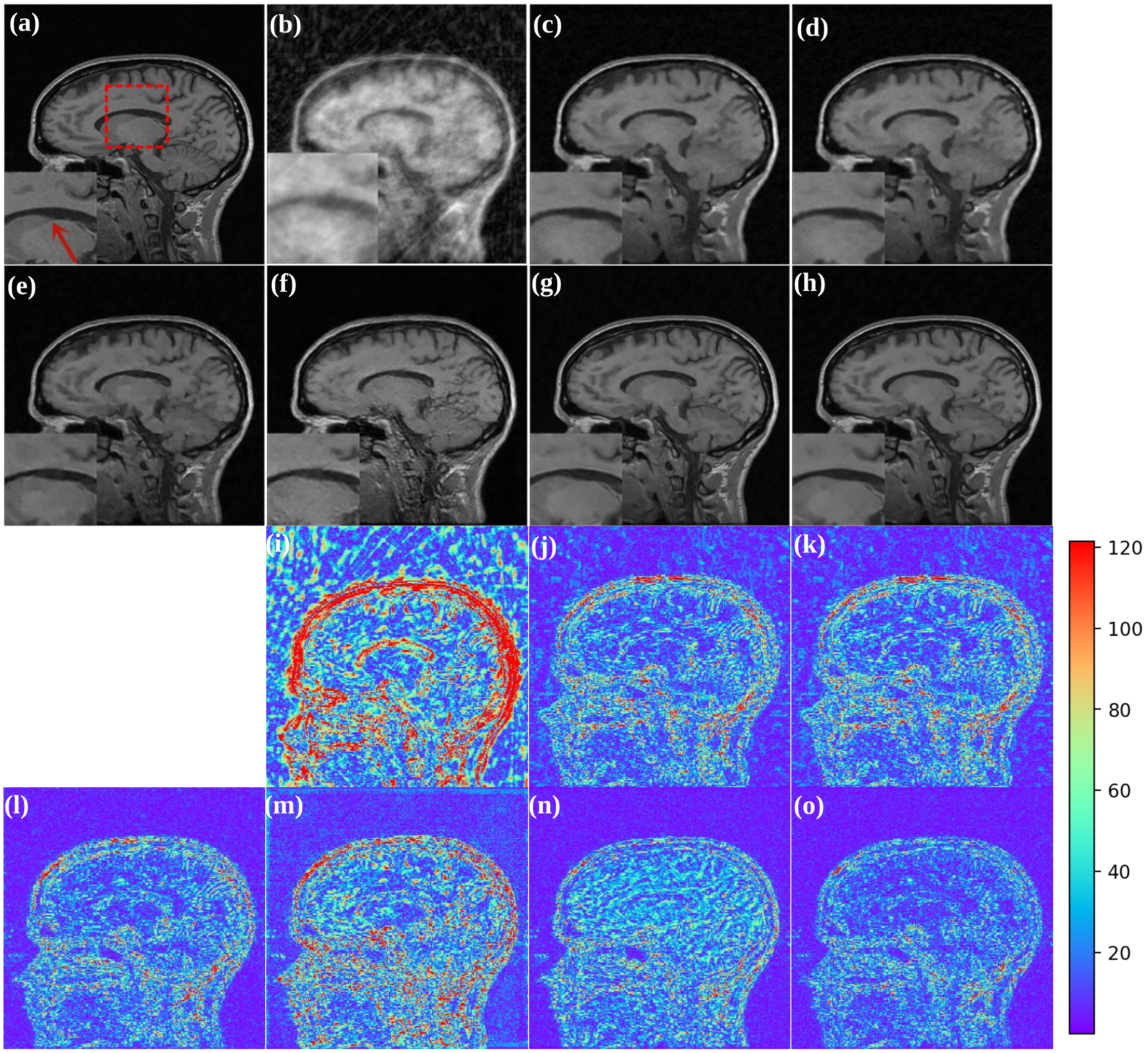

**Fig. 6** Representative visual results with 10% radial sampling rates of (a)full-sampling, (b) zero-filling, (c) PANO (d) DLMRI, (e) ADMM-CSNet, (f) DAGAN, (g) DIMENSION, (h) MD-Recon-Net , and the corresponding error maps of (i) zero-filling, (j) PANO (k) DLMRI, (l) ADMM-CSNet, (m) DAGAN, (n) DIMENSION, (o) MD-Recon-Net.

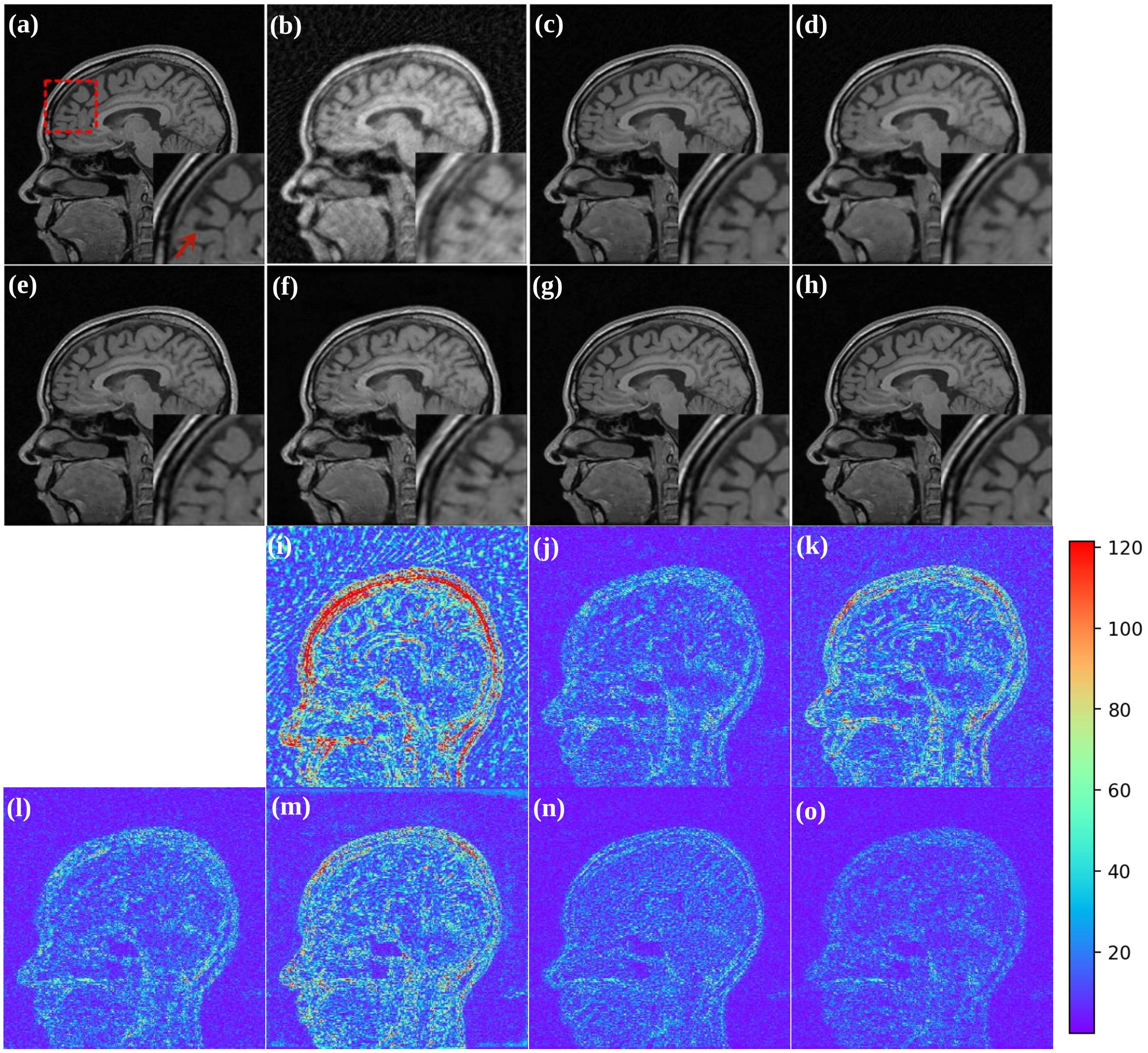

**Fig. 7** Representative visual results with 20% radial sampling rates of (a)full-sampling, (b) zero-filling, (c) PANO (d) DLMRI, (e) ADMM-CSNet, (f) DAGAN, (g) DIMENSION, (h) MD-Recon-Net , and the corresponding error maps of (i) zero-filling, (j) PANO (k) DLMRI, (l) ADMM-CSNet, (m) DAGAN, (n) DIMEN SION, (o) MD-Recon-Net.

**Table III**
Quantitative results associated with different methods' outputs in Figs. 6-8

| | 10% | | 20% | | 25% | |
|---|---|---|---|---|---|---|
| | PSNR | SSIM | PSNR | SSIM | PSNR | SSIM |
| Zero-filling | 23.4653 | 0.5674 | 26.776 | 0.7136 | 28.7171 | 0.7898 |
| PANO | 29.0157 | 0.8177 | 34.7382 | 0.9365 | 36.0017 | 0.9458 |
| DLMRI | 29.5289 | 0.8287 | 31.2278 | 0.8754 | 32.5061 | 0.8935 |
| ADMM-CSNet | 29.9708 | 0.8651 | 34.5634 | 0.9360 | 37.5748 | 0.9606 |
| DAGAN | 26.8826 | 0.7584 | 30.8382 | 0.8714 | 31.9847 | 0.8877 |
| DIMENSION | 30.2237 | 0.8939 | 36.1309 | 0.9595 | 37.757 | 0.9659 |
| MD-Recon-Net | **31.9461** | **0.9028** | **37.854** | **0.9638** | **38.7059** | **0.9673** |

**Table IV**
Quantitative results associated with different methods' outputs in Figs. 9-11

| | Transverse | | Sagittal | | Coronal | |
|---|---|---|---|---|---|---|
| | PSNR | SSIM | PSNR | SSIM | PSNR | SSIM |
| Zero-filling | 27.8661 | 0.8339 | 27.6309 | 0.7391 | 26.3105 | 0.8153 |
| PANO | 37.4555 | 0.9646 | 35.9548 | 0.485 | 35.4816 | 0.9668 |
| DLMRI | 32.8510 | 0.9302 | 32.1459 | 0.8922 | 31.2803 | 0.9270 |
| ADMM-CSNet | 36.9955 | 0.9594 | 35.5754 | 0.9454 | 35.1899 | 0.9649 |
| DAGAN | 32.6258 | 0.9184 | 31.6453 | 0.8821 | 30.7876 | 0.9782 |
| DIMENSION | 39.7598 | 0.9753 | 38.0180 | 0.9652 | 38.0626 | 0.9782 |
| MD-Recon-Net | **40.3056** | **0.9770** | **38.8691** | **0.9692** | **38.8463** | **0.9802** |

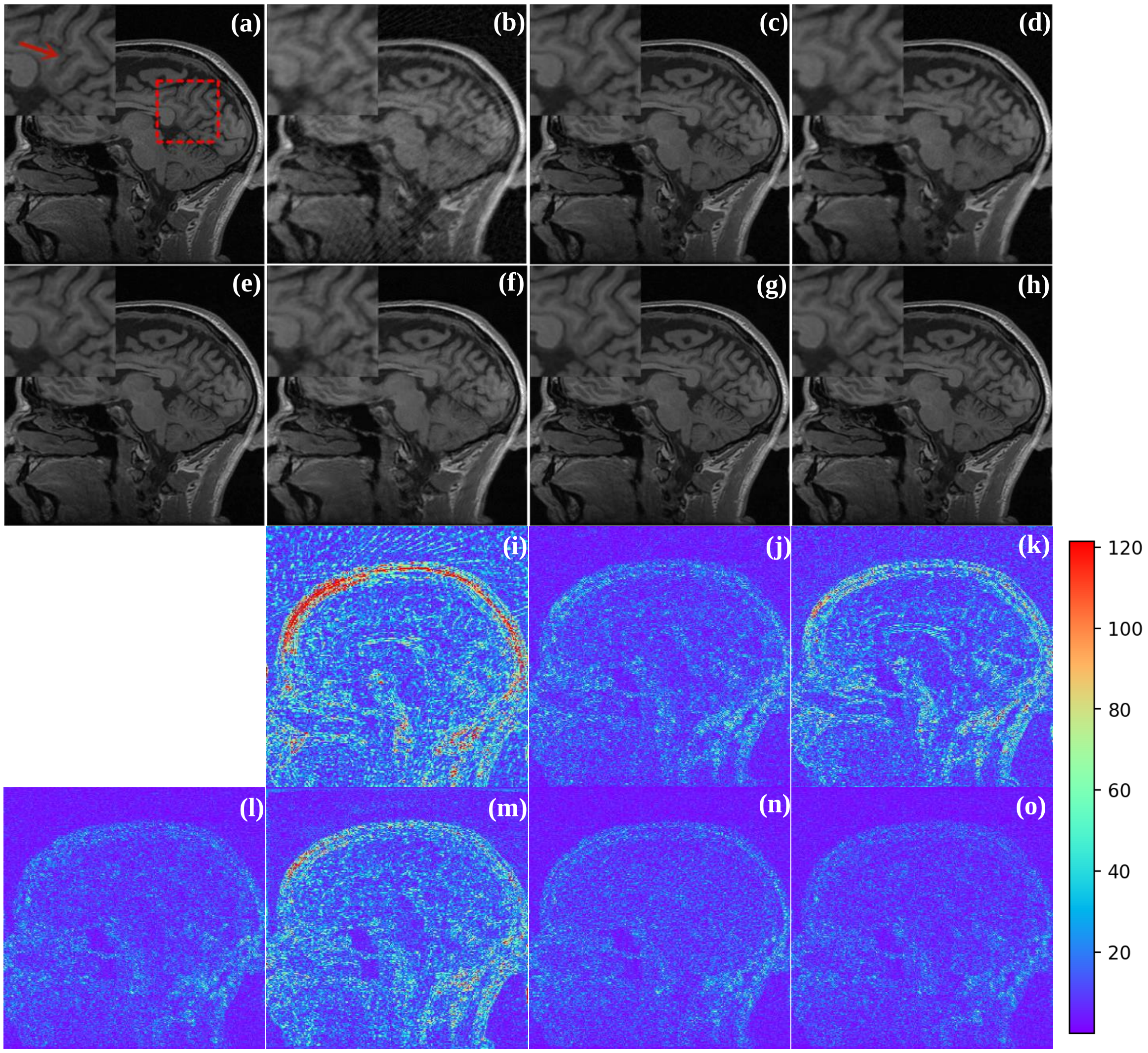

**Fig. 8** Representative visual results with 25% radial sampling rates of (a)full-sampling, (b) zero-filling, (c) PANO (d) DLMRI, (e) ADMM-CSNet, (f) DAGAN, (g) DIMENSION, (h) MD-Recon-Net , and the corresponding error maps of (i) zero-filling, (j) PANO (k) DLMRI, (l) ADMM-CSNet, (m) DAGAN, (n) DIMENSION, (o) MD-Recon-Net.

$$\mathrm{PSNR} = 20\, log_{10} \frac{255}{\sqrt{\frac{1}{mn}\sum_{i=0}^{m-1}\sum_{j=0}^{n-1}[I(i,j) - K(i,j)]^2}} \quad (7)$$

$$\mathrm{SSIM} = \frac{(2\mu_x\mu_{\hat{x}} + c_1)(2\sigma_{x\hat{x}} + c_2)}{(\mu_x^2 + \mu_{\hat{x}}^2 + c_1)(\sigma_x^2 + \sigma_{\hat{x}}^2 + c_2)} \quad (8)$$

To qualitatively evaluate the results, reader studies similar to the qualitative metric in [44] was performed by two radiologists with 6 and 8 years of clinical experience respectively. They assessed these images independently and blindly to provide their scores. The ground truth images were treated as the golden standard. For each testing set of images, 30 slices were randomly selected and the scores were reported as means ±SDs (average scores of two radiologists ± standard deviations). Overall quality (OQ) was adopted as subjective metric on the five-point scale (1 = unacceptable and 5 = excellent).

### *B. Single vs Dual Domain*

First, we evaluated the effectiveness of the dual-domain method by comparing the proposed MD-Recon-Net with its two variants, MD-Recon-Net-k and MD-Recon-Net-s. The architectures of the two variants are similar to that of MD-Recon-Net, but the processed data vary. MD-Recon-Net-k only deals with the undersampled *k*-space data, so only KDC modules are maintained, and only one IFT module is used at the end of the network. In contrast to MD-Recon-Net-k, MD-Recon-Net-s only operates on the spatial data, so only the SDC module is preserved, and the IFT module is used at the beginning of the model.

Table I summarizes the statistical quantitative results of the three methods with different sampling rates and strategies. MD-Recon-Net and MD-Recon-Net-s significantly outperform MD-Recon-Net-k in all cases, leading to a consistent conclusion with [38] that *k*-space based methods usually have limited performance because of the complex relationship between *k*-space data with reconstructed images. MD-Recon-Net achieves better scores than MD-Recon-Net-s, and when the sampling rate is lower in the radial sampling case, the improvement is more obvious. Fig. 5. depicts one representative slice reconstructed by three different models, and the radial sampling mask was employed with a 20% sampling rate. The visual effects in Fig. 5 are coherent with the quantitative results that the MD-Recon-Net and MD-Recon-Net-s methods obtained better results than MD-Recon-Net-k. The error maps show that MD-Recon-Net is able to suppress artifacts better and preserve more details than MD-Recon-Net-s. In summary, our proposed method achieves the best quantitative scores and visual effects, which can be treated as powerful evidence that dual-domain reconstruction is superior to single-domain reconstruction.

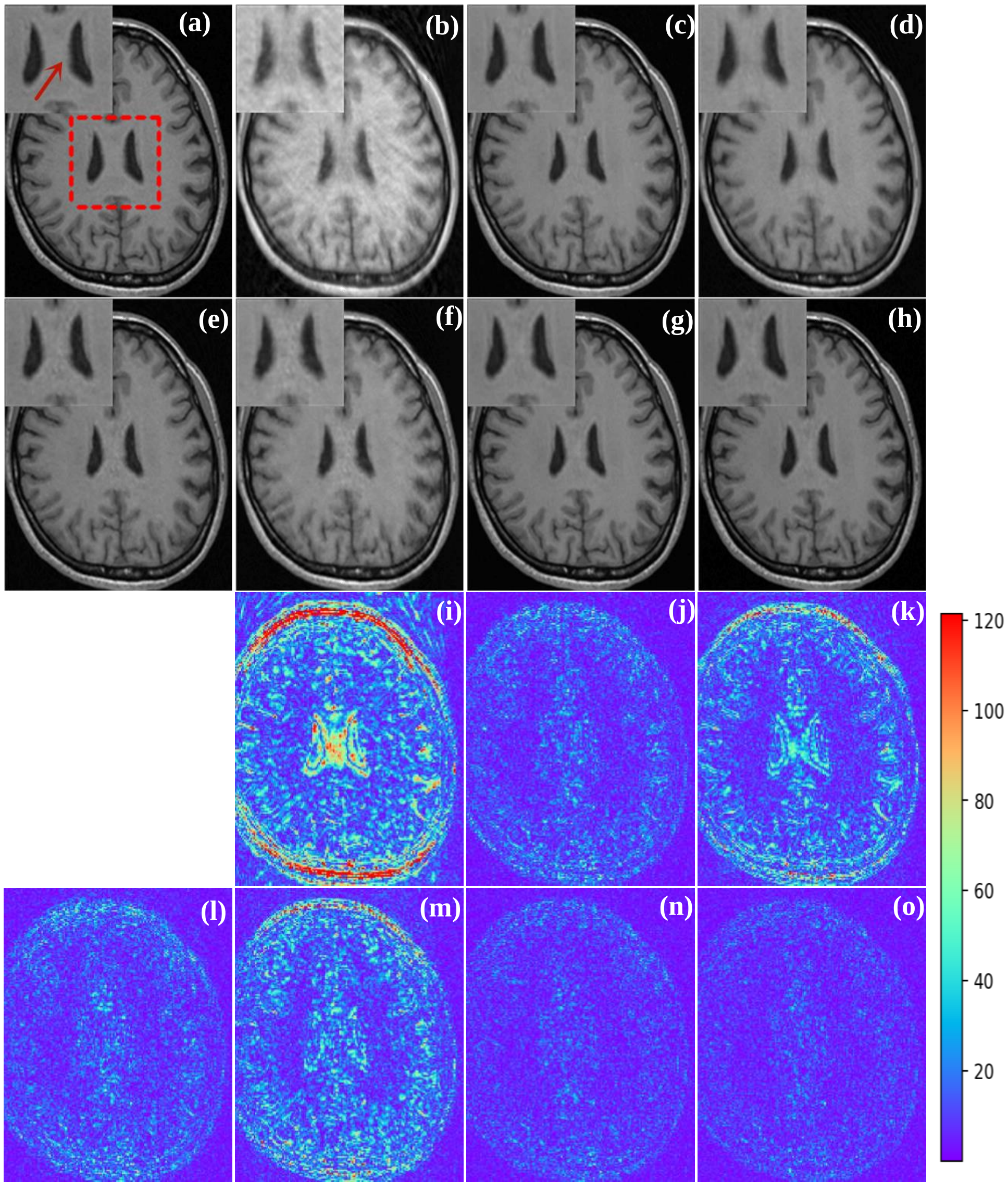

**Fig. 9** Representative visual results along traverse direction with 20% radial sampling mask of (a)full-sampling, (b) zero-filling, (c) PANO (d) DLMRI, (e) ADMM-CSNet, (f) DAGAN, (g) DIMENSION, (h) MD-Recon-Net , and the corresponding error maps of (i) zero-filling, (j) PANO (k) DLMRI, (l) ADMM-CSNet, (m) DAGAN, (n) DIMENSION, (o) MD-Recon-Net.

### C. Comparison with Other State-of-the-art Methods

In this section, to demonstrate the performance of the proposed MD-Recon-Net, five state-of-the-art CS-MRI methods that belong to different groups (PANO [43], DLMRI [21], ADMM-CSNet [30], DAGAN [25], and DIMENSION [36]) were compared. Table II shows the statistical quantitative results of different methods with different sampling rates and strategies on the test set. The best results for each measurement are bolded. Generally, DL-based methods had better performance than traditional CS-based methods in terms of both metrics (except DAGAN). A possible reason is that DAGAN contains downsampling operations, which may lead to loss of detail. Unfortunately, some streak-like artifacts have been proven difficult to remove efficiently by post-processing methods [45]. We also observed that dual-domain methods (DIMENSION and MD-Recon-Net) outperformed single-domain methods (ADMM-CSNet and DAGAN). The introduction of the *k*-space data constraint gave ADMM-CSNet better performance than DAGAN. Our method obtained the best scores in both metrics and had the smallest variations in all cases. Meanwhile, the reader study confirmed the visual and quantitative results that our method gained the highest means and relatively small variation in each case. Further, once the networks are trained, the running times for DL-based methods are much shorter than those of CS-based methods. Table II shows the average execution times of different methods and MD-Recon-Net obtained the fastest execution speed, which implicitly proves the smaller scale of model size.

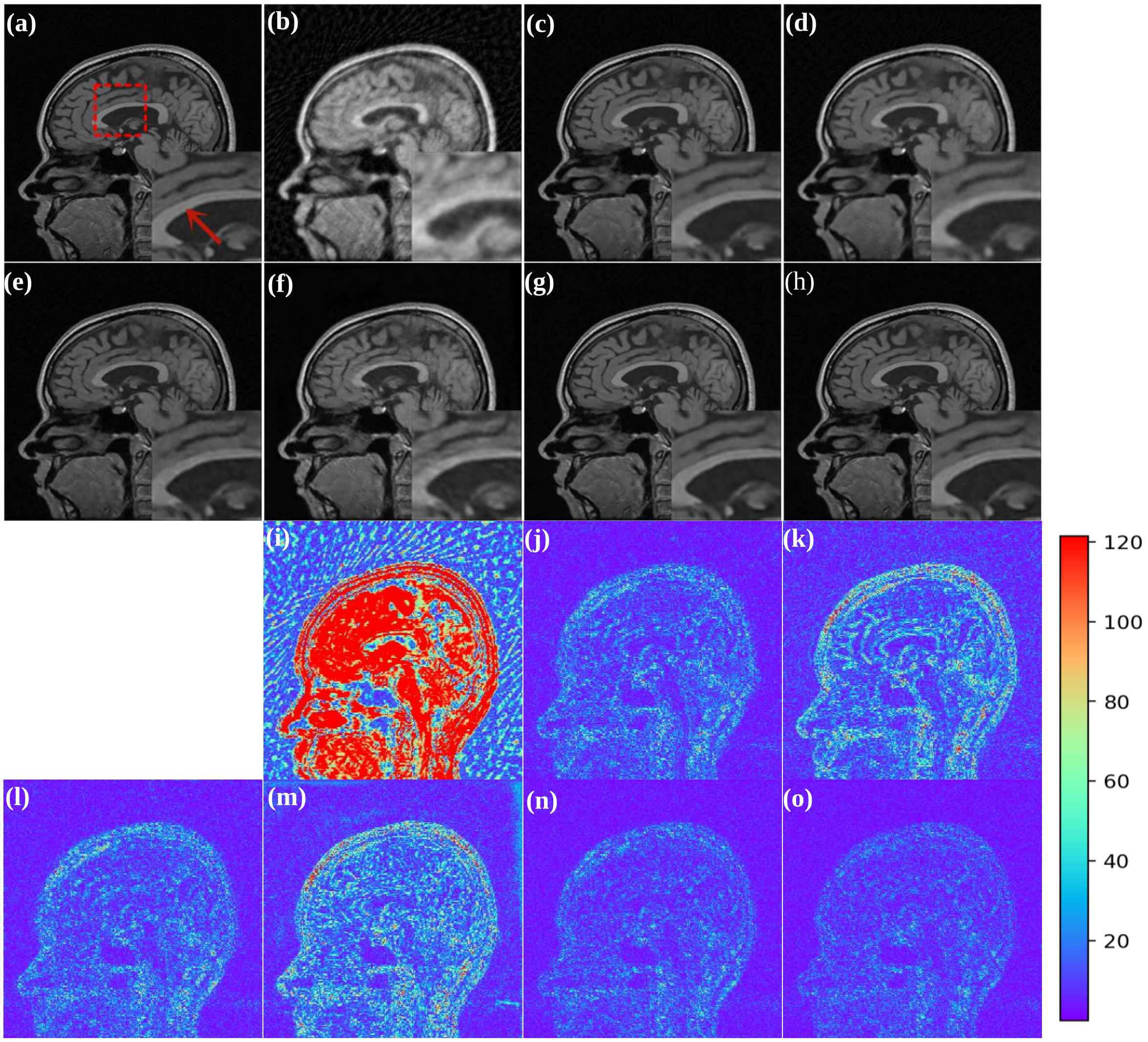

**Fig. 10** Representative visual results along sagittal direction with 20% radial sampling mask of (a)full-sampling, (b) zero-filling, (c) PANO (d) DLMRI, (e) ADMM-CSNet, (f) DAGAN, (g) DIMENSION, (h) MD-Recon-Net , and the corresponding error maps of (i) zero-filling, (j) PANO (k) DLMRI, (l) ADMM-CSNet, (m) DAGAN, (n) DIMENSION, (o) MD-Recon-Net.

### 1) *Different Sampling Rates*

Representative visual results reconstructed by different methods with different radial sampling rates (10%, 20%, and 25%) are shown in Figs. 6-8. Figs. 6-8 demonstrate the results at 10%, 20%, and 25% sampling rates, respectively. For each sampling rate, the first row contains the reconstructed images (ground truth, zero-filling, and different reconstruction methods), and the second row shows the corresponding error maps compared with the ground truth. Figs. 6-8 shows that all methods are able to suppress artifacts to varying degrees. When the sampling rate is high, the visual differences among different results are inconspicuous. Some differences can be recognized by the error maps: PANO, ADMM-CSNet, DIMENSION, and MD-Recon-Net had less error than the other two methods. At lower sampling rates, the merits of dual-domain methods become noticeable. For the 10% sampling rate case, some details near the parietooccipital sulcus are blurred, except on the DIMENSION and MD-Recon-Net results. The details appear more clearly in the MD-Recon-Net results, and the error maps also support this observation. Table III gives the quantitative results associated with the reconstructed images in Figs. 6-8, which are consistent with the visual effects. In the magnified parts in Fig. 6-8, it is clear that the DL-based methods can yield more details and the outputs of the classical methods are oversmooth. From the area marked by the red arrows, the performances of MD-Recon-Net can be better demonstrated than the other methods.

Figs. 9-11 shows the reconstructed results along traverse sagittal and coronal directions with 20% radial sampling respectively. For the results of DL-based methods in Figs. 9-11, only one network was trained with the undersampled sagittal data instead of training three different models for three different panels. All the methods were applied on the sagittal data. After that, we stacked all the results of sagittal planes to a complete volume. Then the results of other two planes can be simply obtained along traverse and coronal directions from the 3D volume data. Meanwhile, some regions, which can visually differentiate the ability of detail recovery of different methods, are magnified and some noticeable structures or details were indicate by red arrows. Table IV gives the associated quantitative results. The results in Figs. 9-11 demonstrate a similar trend to that in Figs. 6-8: two dual-domain methods

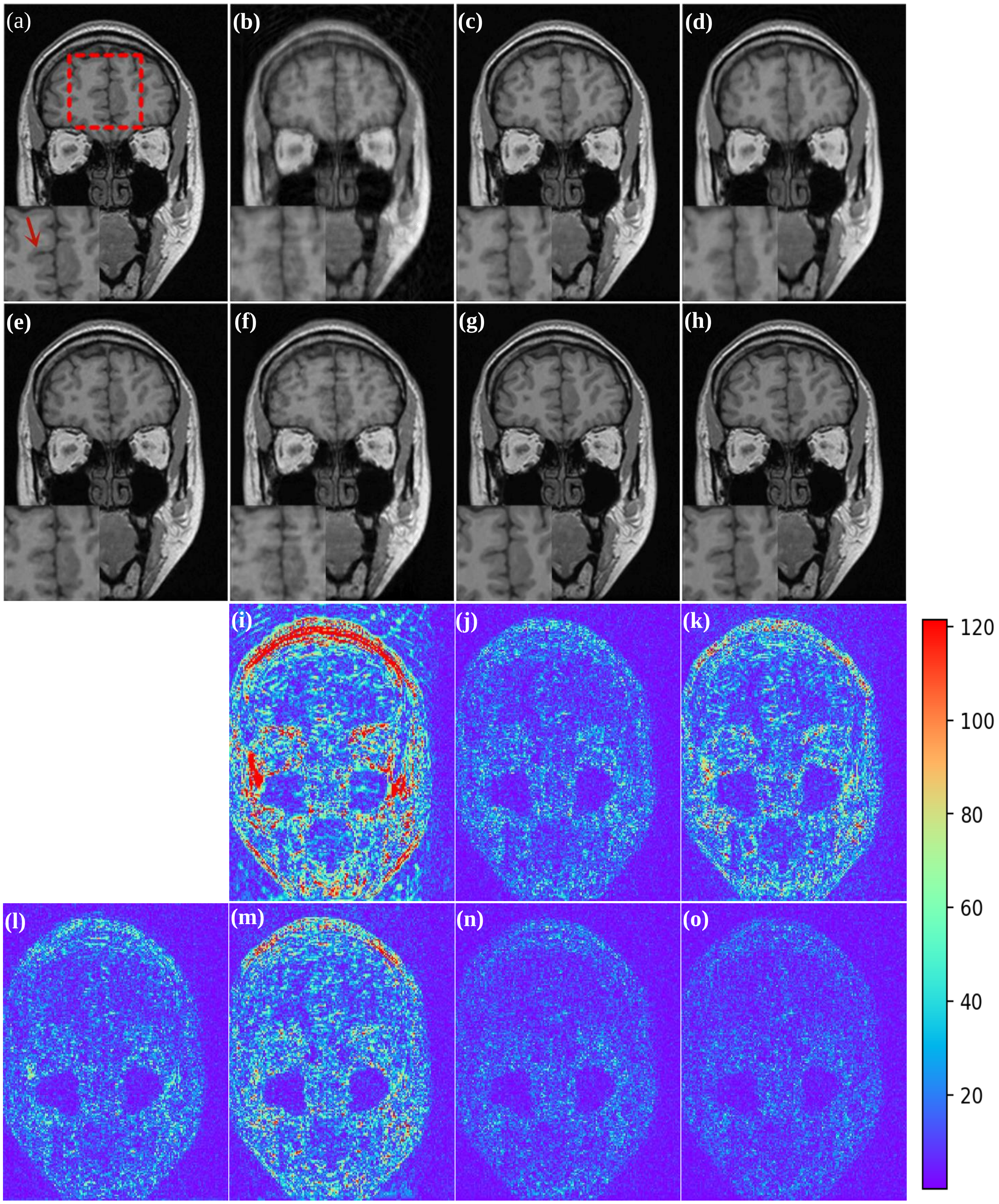

**Fig. 11** Representative visual results along coronal direction with 20% radial sampling mask of (a)full-sampling, (b) zero-filling, (c) PANO (d) DLMRI, (e) ADMM-CSNet, (f) DAGAN, (g) DIMENSION, (h) MD-Recon-Net , and the corresponding error maps of (i) zero-filling, (j) PANO (k) DLMRI, (l) ADMM-CSNet, (m) DAGAN, (n) DIMENSION, (o) MD-Recon-Net.

**Table V**

Quantitative results associated with different methods' outputs in Figs. 12-14

| | Radial | | Cartesian | | Random | |
|---|---|---|---|---|---|---|
| | PSNR | SSIM | PSNR | SSIM | PSNR | SSIM |
| Zero-filling | 27.9772 | 0.7444 | 25.1022 | 0.6385 | 25.3382 | 0.6474 |
| DLMRI | 31.5036 | 0.8613 | 27.2289 | 0.7214 | 30.1008 | 0.8250 |
| PANO | 34.5600 | 0.9251 | 28.3465 | 0.7548 | 33.8246 | 0.9088 |
| ADMM-CSNet | 34.3513 | 0.9239 | 29.3620 | 0.7952 | 34.7296 | 0.9282 |
| DAGAN | 31.4096 | 0.8637 | 29.0897 | 0.7837 | 28.2628 | 0.7908 |
| DIMENSION | 35.9473 | 0.9508 | 30.0727 | 0.8282 | 36.0129 | 0.9466 |
| MD-Recon-Net | **37.3059** | **0.9548** | **30.4447** | **0.8333** | **37.6436** | **0.9566** |

**Table VI**

Statistical quantitative results (PSNR, SSIM and OQ) of different methods for the knee test set

| | PSNR | SSIM | OQ |
|---|---|---|---|
| Zero-filling | 29.14±1.59 | 0.72±0.05 | 1.13±0.16 |
| PANO | 30.48±1.60 | 0.75±0.04 | 2.95±0.52 |
| DLMRI | 30.43±1.49 | 0.75±0.04 | 2.93±0.88 |
| ADMM-CSNet | 30.58±1.63 | 0.75±0.04 | 3.11±0.56 |
| DAGAN | 30.39±1.55 | **0.76±0.04** | 2.47±0.33 |
| DIMENSION | 30.43±1.73 | 0.74±0.04 | 3.22±0.47 |
| MD-Recon-Net | **30.81±1.70** | **0.76±0.04** | **3.35±0.52** |

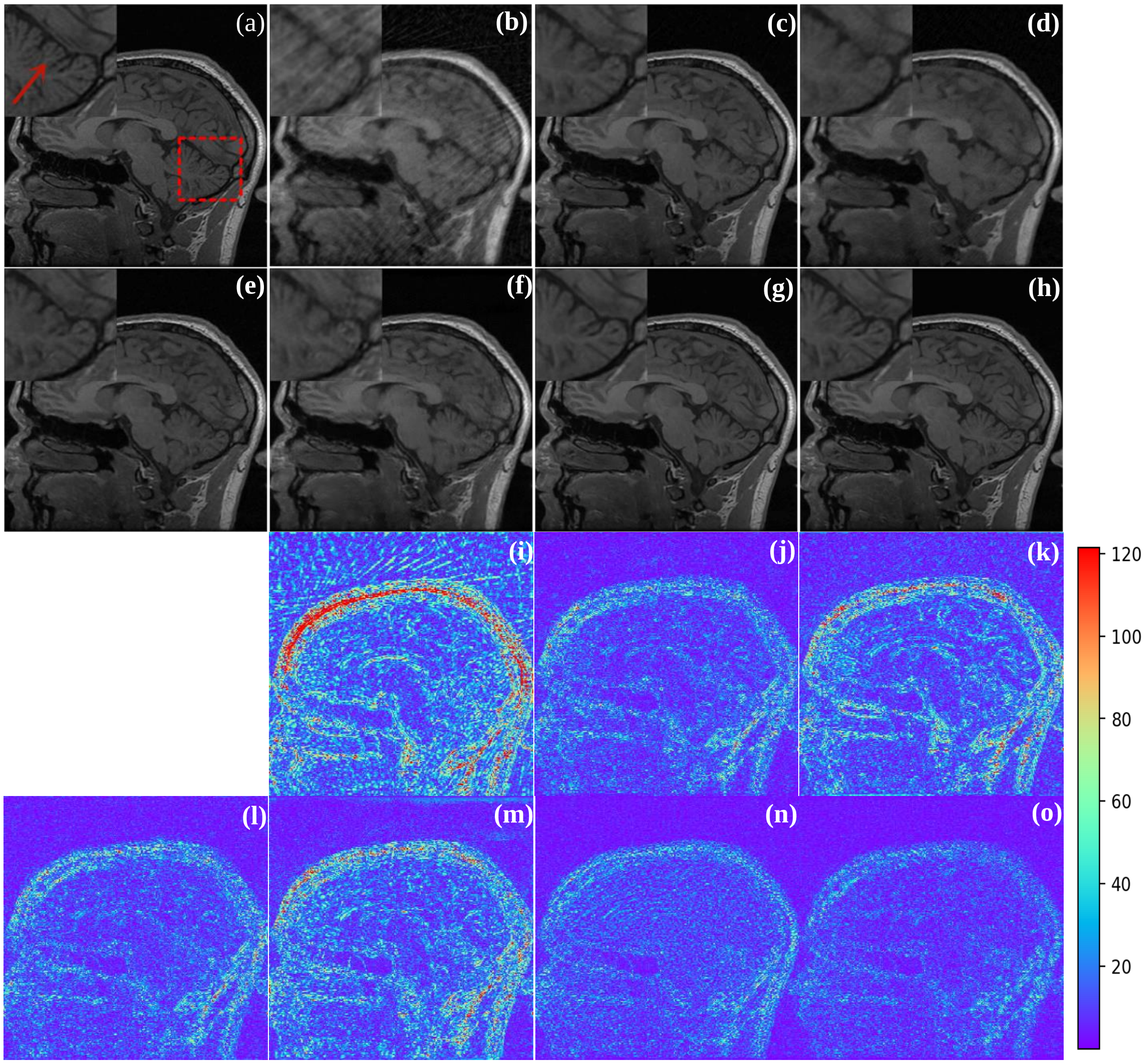

**Fig. 12** Representative visual results with 20% radial sampling rate of (a)full-sampling, (b) zero-filling, (c) PANO (d) DLMRI, (e) ADMM-CSNet, (f) DAGAN, (g) DIMENSION, (h) MD-Recon-Net , and the corresponding error maps of (i) zero-filling, (j) PANO (k) DLMRI, (l) ADMM-CSNet, (m) DAGAN, (n) DIMENSION, (o) MD-Recon-Net.

show better performance in both qualitative and quantitative aspects than other methods.

*2) Different Sampling Strategies*

In this subsection, different sampling strategies including radial, Cartesian, and Gaussian random sampling masks were considered to validate the performance of the proposed MD-Recon-Net. The sampling rates of all sampling methods were uniformly set to 20%. Figs. 12-14 shows the reconstructed results of different methods with different sampling masks. Figs. 12-14 shows that all methods can eliminate most of the aliasing artifacts, and MD-Recon-Net demonstrates competitive abilities in both artifact removal and detail preservation. In all cases, MD-Recon-Net obtained the best resolution and details, and the magnified regions and error maps also confirm this observation. Especially, the results for Cartesian sampling were worse than those of other sampling strategies. Because the low frequency of the centered Fourier spectrum (*k*-space) contains the major content of the spectrum, the Cartesian sampling strategy collects less information from this part and produces more severe artifacts. The PANO and DLMRI results still contain some artifacts, and the structural details are blurred heavily. ADMM-CSNet and DAGAN suppress the artifacts better, but some sulcus structures are blurred. DIMENSION and MD-Recon-Net achieved similar performance in this regard and maintained more details than other methods. Table V provides the quantitative results for the images in Figs. 12-14, and the proposed MD-Recon-Net outperforms all the other methods in terms of both metrics.

## D. *Other Issues*

*1) Robustness*

To verify the generalizability of our proposed method, we also tested our model using the knee dataset[6]. The raw data were acquired using the 3D fast-spin-echo (FSE) sequence with proton density weighting including fat saturation comparison on a 3.0T whole body MR system (Discovery MR 750, DV22.0,

[6] http://mridata.org/

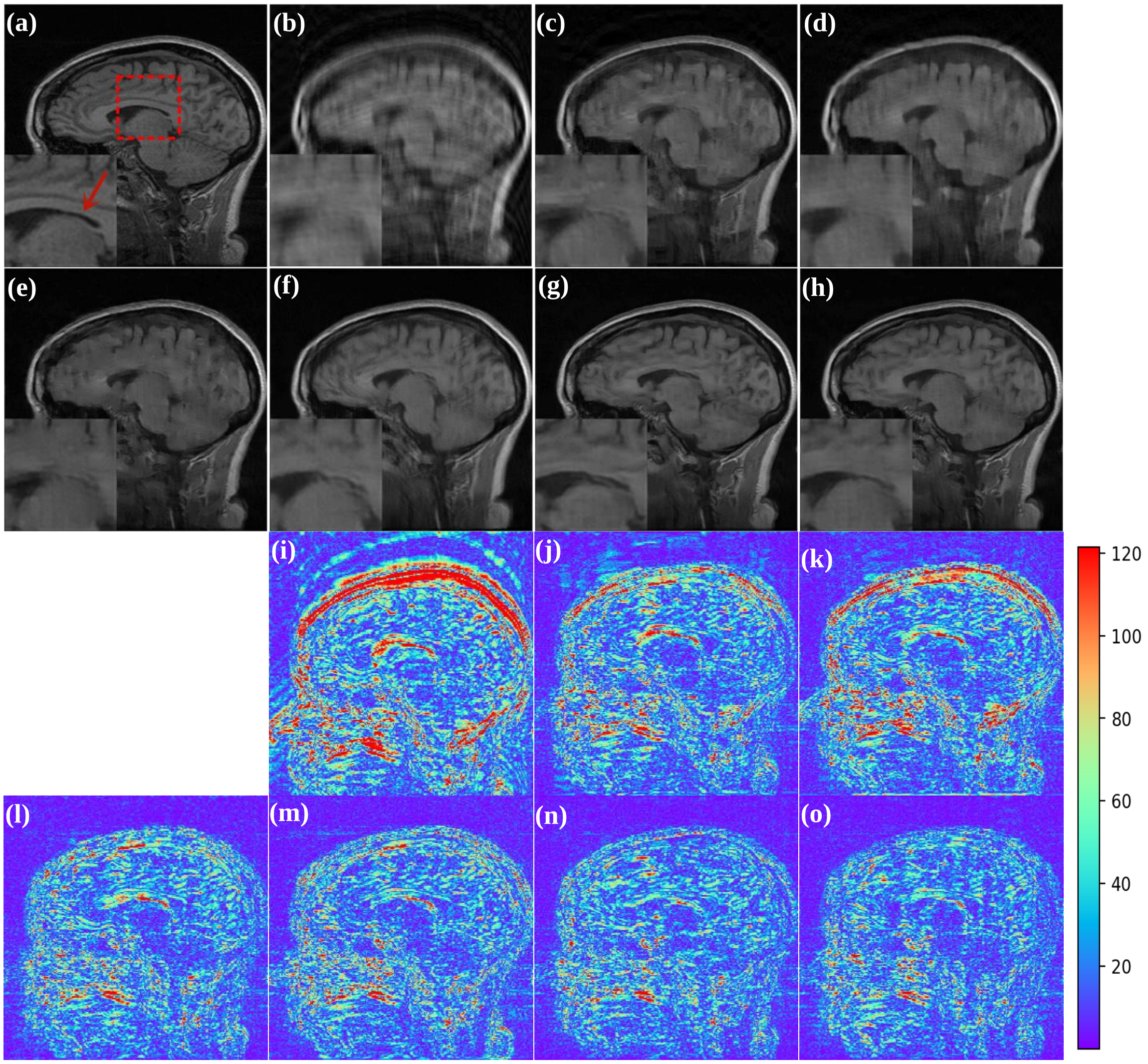

**Fig. 13** Representative visual results with 20% cartesian sampling rate of (a)full-sampling, (b) zero-filling, (c) PANO (d) DLMRI, (e) ADMM-CSNet, (f) DAGAN, (g) DIMENSION, (h) MD-Recon-Net , and the corresponding error maps of (i) zero-filling, (j) PANO (k) DLMRI, (l) ADMM-CSNet, (m) DAGAN, (n) DIMENSION, (o) MD-Recon-Net.

GE Healthcare, Milwaukee, WI, USA). The repetition time and echo time were 1,550 ms and 25 ms, respectively. There were 256 slices in total, and each slice's thickness was 0.6 mm. The field of view (FOV) was defined as $160\times160$ mm$^2$, and the size of acquisition matrix was $320\times320$. The voxel size was 0.5 mm, and the number of coils was 8. However, our proposed method is based on single coil data, so we used a coil compression algorithm[7] to produce single coil *k*-space data. Table VI shows the statistical quantitative results of different methods with 20% sampling rate, and with Cartesian sampling strategy on the whole set. It can be seen that our MD-Recon-Net obtained the best scores in both metrics. The results of reader study in Table VI are coherent with the quantitative results that although the differences among different methods are smaller than the ones in previous dataset, the proposed MD-Recon-Net still performed best in all cases. Fig. 15 shows a representative slice reconstructed by different methods. DIMENSION and MD-Recon-Net had the best visual effects. The results from other methods appear slightly blurred, even if the quantitative scores are close.

**Table VII**
Complexity evaluation of different network models

| | Parameters | FLOPs(M) |
|---|---|---|
| DAGAN | 98600449 | **2001** |
| DIMENSION | 848010 | 34577 |
| MD-Recon-Net | **289319** | 2106 |

### 2) *Model Scale and FLOPs*

The complexity of the network model is an important factor to evaluate the practicability of networks, because it has an important impact on training difficulty and memory requirements. Thus, to be fair, we need to compare model complexity between the available options. In general, the complexity of deep learning is measured by two metrics: the amount of parameters (model scale) and floating point

[7] http://mrsrl.stanford.edu/~tao/software.html

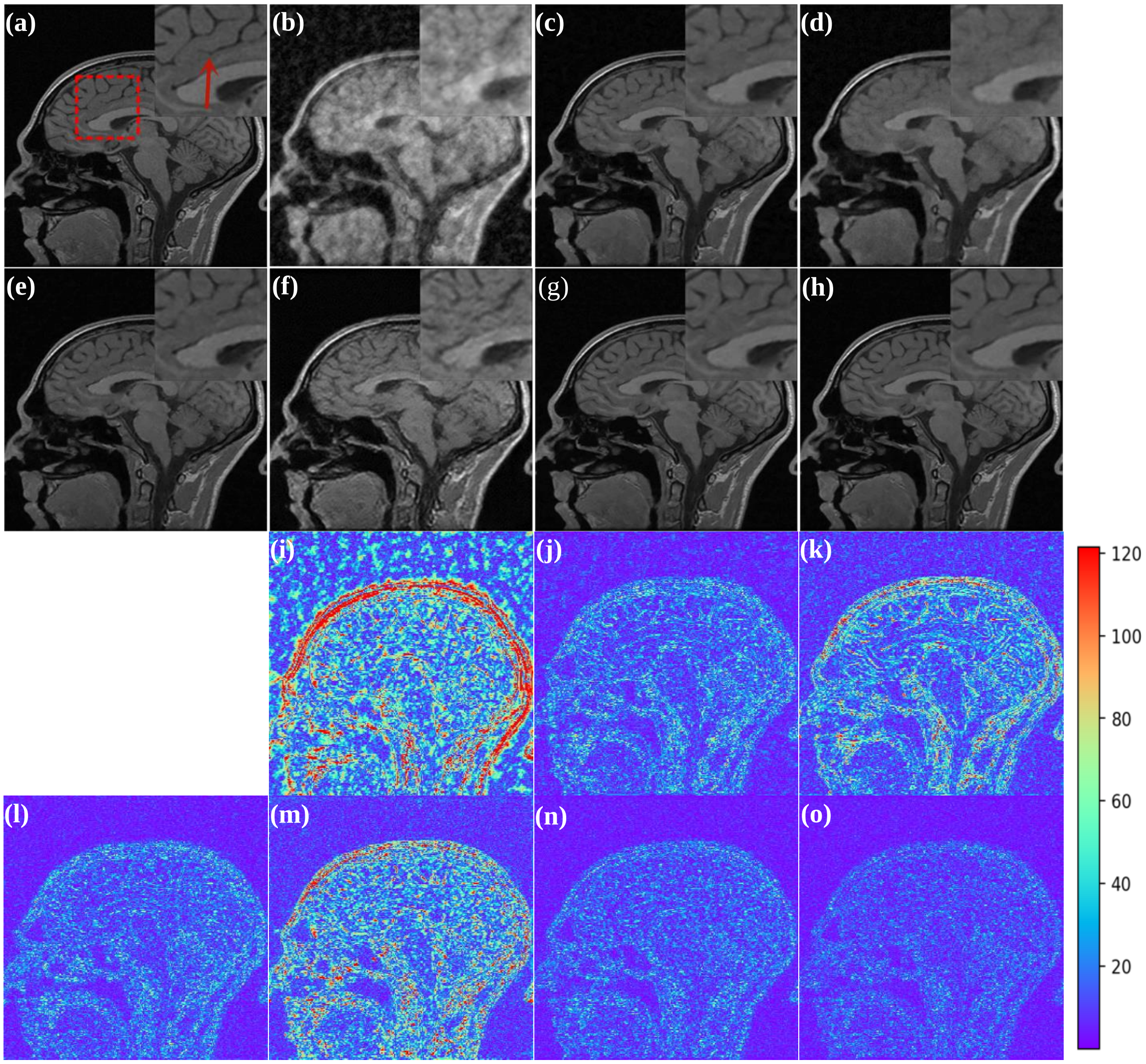

**Fig. 14** Representative visual results with 20% gaussian random sampling rate of (a)full-sampling, (b) zero-filling, (c) PANO (d) DLMRI, (e) ADMM-CSNet, (f) DAGAN, (g) DIMENSION, (h) MD-Recon-Net , and the corresponding error maps of (i) zero-filling, (j) PANO (k) DLMRI, (l) ADMM-CSNet, (m) DAGAN, (n) DIMENSION, (o) MD-Recon-Net.

operations(FLOPs). The three DL-based methods (DAGAN, DIMENSION, and MD-Recon-Net) were compared and the results are shown in Table VII, which shows that the amount of parameters in MD-Recon-Net is much smaller than that of the other methods. In detail, the parameter amount of DIMENSION is 2.93 times that of MD-Recon-Net, and DAGAN has 340.8 times that of MD-Recon-Net. In terms of FLOPs, DAGAN achieved the best score, but MD-Recon-Net's score was very close. This result is not surprising, as DAGAN contains some large-scale convolutional kernels (e.g., 4×4), and the numbers of filters in some layers are relatively large (e.g., 512 and 1,024). The parameter quantity of DAGAN is very large, but downsampling operations (from 256×256 to 1×1) also exist in DAGAN, which make its number of FLOPs drop sharply. DIMENSION's number of FLOPs is mainly caused by its usage of 3-D complex convolution [36, 46], which consists of a 3-D convolution operation. In general, our proposed MD-Recon-Net has lower model complexity than the other two methods have.

## IV. CONCLUSION

In this paper, we proposed a novel cascaded dual-domain CNN for fast MRI. Different from existing dual-domain methods, our proposed MD-Recon-Net deals with the *k*-space and spatial data simultaneously. In addition, data fusion modules exchange the intermediate results from different domains. Meanwhile, data consistency modules in both *k*-space and the spatial domain can further improve performance.

The experimental results demonstrate that using dual-domain information improves both the proposed network's qualitative and quantitative aspects. Compared with several state-of-the-art methods (PANO, DLMRI, ADMM-CSNet, DAGAN, and DIMENSION), our proposed MD-Recon-Net can effectively remove artifacts while preserving more detail at different sampling rates (10%, 20%, and 25%) with different sampling strategies (radial, Cartesian, and Gaussian random sampling). The proposed model's complexity was also analyzed, and the proposed MD-Recon-Net has low model scale and competitive FLOPS, which means that our method not only has competitive performance compared with the state-of-the-art methods, but also requires less computational resources for training and testing.

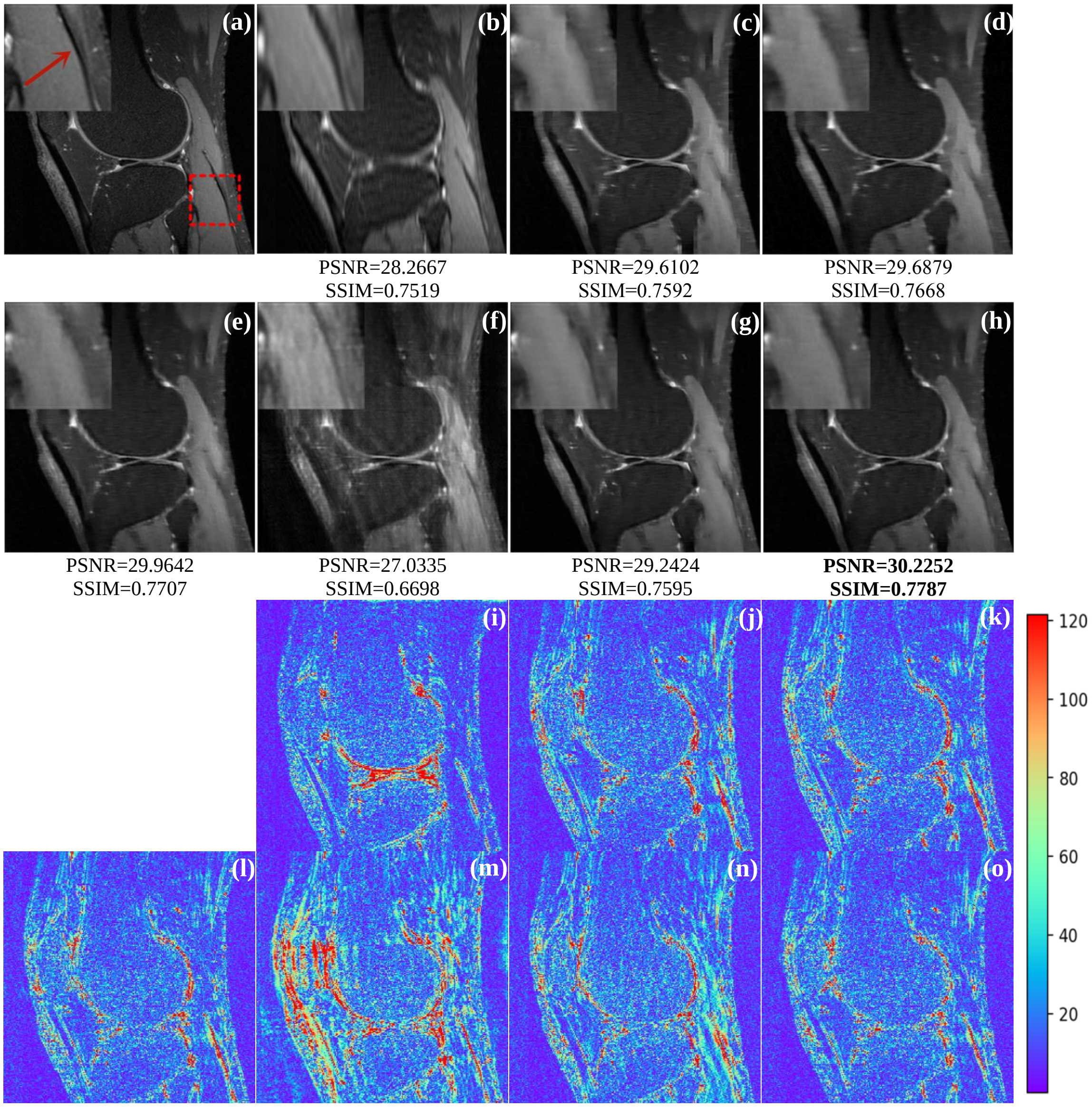

**Fig. 15** Representative visual results on knee data with 20% Cartesian sampling rate of (a)full-sampling, (b) zero-filling, (c) PANO (d) DLMRI, (e) ADMM-CSNet, (f) DAGAN, (g) DIMENSION, (h) MD-Recon-Net , and the corresponding error maps of (i) zero-filling, (j) PANO (k) DLMRI, (l) ADMM-CSNet, (m) DAGAN, (n) DIMENSION, (o) MD-Recon-Net.

It also must be mentioned that although our experiments used real data, the sampling patterns we used were simulated, which may be different from the realistic scans. The possible solution is that we can train a network for a specific scanner with its specific sampling pattern. Meanwhile, parallel imaging has been widely used to accelerate MRI scan. The technique requires the utilization of multiple physical receiver coils to simultaneously record different views of the object being imaged. Parallel imaging is the default option for many scan protocols and it is supported by almost all modern clinical MRI scanners. Our proposed model cannot be applied directly for the parallel imaging, but some tricks may help. For examples, coil compression algorithms can be used at first to combine the multiple coil k-space data or a coil data fusion layer can be added to current network.

In our future work, we will try to solve the issues mentioned above and more datasets and sampling strategies will be included for validation. In addition, some advanced techniques such as attention mechanisms and more complex loss functions will be considered.

## V. Acknowledgments

We thank Richard Lipkin, PhD, from Liwen Bianji, Edanz Group China (www.liwenbianji.cn/ac), for editing the English text of a draft of this manuscript.

## References

[1] J. Bland, M. A. Belzunce, S. Ellis, C. J. McGinnity, A. Hammers, and A. J. Reader, "Spatially compact MR-Guided kernel EM for PET image reconstruction," *IEEE Trans. Radiat. Plasma Med. Sci.*, vol. 2, no. 5, pp. 470-482, 2018.

[2] J. Bland *et al.*, "MR-guided kernel EM reconstruction for reduced dose PET imaging," *IEEE Trans. Radiat. Plasma Med. Sci.*, vol. 2, no. 3, pp. 235-243, 2017.

[3] K. M. Ropella-Panagis, N. Seiberlich, and V. Gulani, "Magnetic resonance fingerprinting: Implications and opportunities for PET/MR," *IEEE Trans. Radiat. Plasma Med. Sci.*, vol. 3, no. 4, pp. 388-399, 2019.

[4] J. Hamilton, D. Franson, and N. Seiberlich, "Recent advances in parallel imaging for MRI," *Prog. Nucl. Magn. Reson. Spectrosc.*, vol. 101, pp. 71-95, 2017.

[5] P. B. Roemer, W. A. Edelstein, C. E. Hayes, S. P. Souza, and O. M. Mueller, "The NMR phased array," *Magn. Reson. Med.*, vol. 16, no. 2, pp. 192-225, 1990.

[6] K. P. Pruessmann, M. Weiger, P. Börnert, and P. Boesiger, "Advances in sensitivity encoding with arbitrary k-space trajectories," *Magn. Reson. Med.*, vol. 46, no. 4, pp. 638-651, 2001.

[7] K. P. Pruessmann, M. Weiger, M. B. Scheidegger, and P. Boesiger, "SENSE: sensitivity encoding for fast MRI," *Magn. Reson. Med.*, vol. 42, no. 5, pp. 952-962, 1999.

[8] D. Liang, B. Liu, J. Wang, and L. Ying, "Accelerating SENSE using compressed sensing," *Magn. Reson. Med.*, vol. 62, no. 6, pp. 1574-1584, 2009.

[9] P. Kellman, F. H. Epstein, and E. R. McVeigh, "Adaptive sensitivity encoding incorporating temporal filtering (TSENSE)," *Magn. Reson. Med.*, vol. 45, no. 5, pp. 846-852, 2001.

[10] Y. J. Ma, W. Liu, X. Tang, and J. H. Gao, "Improved SENSE imaging using accurate coil sensitivity maps generated by a global magnitude-phase fitting method," *Magn. Reson. Med.*, vol. 74, no. 1, pp. 217-224, 2015.

[11] J. Wang, T. Kluge, M. Nittka, V. Jellus, B. Kuehn, and B. Kiefer, "Parallel acquisition techniques with modified SENSE reconstruction mSENSE," in *Proceedings of the First Würzburg Workshop on Parallel Imaging Basics and Clinical Applications*, 2001, p. 89.

[12] C. A. McKenzie, M. A. Ohliger, E. N. Yeh, M. D. Price, and D. K. Sodickson, "Coil-by-coil image reconstruction with SMASH," *Magn. Reson. Med.*, vol. 46, no. 3, pp. 619-623, 2001.

[13] M. Bydder, D. J. Larkman, and J. V. Hajnal, "Generalized SMASH imaging," *Magn. Reson. Med.*, vol. 47, no. 1, pp. 160-170, 2002.

[14] D. K. Sodickson, "Simultaneous acquisition of spatial harmonics (SMASH): ultra-fast imaging with radiofrequency coil arrays," ed: Google Patents, 1999.

[15] M. A. Griswold *et al.*, "Generalized autocalibrating partially parallel acquisitions (GRAPPA)," *Magn. Reson. Med.*, vol. 47, no. 6, pp. 1202-1210, 2002.

[16] W. E. Kyriakos *et al.*, "Sensitivity profiles from an array of coils for encoding and reconstruction in parallel (SPACE RIP)," *Magn. Reson. Med.*, vol. 44, no. 2, pp. 301-308, 2000.

[17] M. Lustig and J. M. Pauly, "SPIRiT: iterative self-consistent parallel imaging reconstruction from arbitrary k-space," *Magn. Reson. Med.*, vol. 64, no. 2, pp. 457-471, 2010.

[18] D. L. Donoho, "Compressed sensing," *IEEE Trans. Inf. Theory*, vol. 52, no. 4, pp. 1289-1306, 2006.

[19] M. Lustig, D. L. Donoho, J. M. Santos, and J. M. Pauly, "Compressed sensing MRI," *IEEE signal processing magazine*, vol. 25, no. 2, p. 72, 2008.

[20] M. Lustig, D. Donoho, and J. M. Pauly, "Sparse MRI: The application of compressed sensing for rapid MR imaging," *Magn. Reson. Med.*, vol. 58, no. 6, pp. 1182-1195, 2007.

[21] S. Ravishankar and Y. Bresler, "MR image reconstruction from highly undersampled k-space data by dictionary learning," *IEEE Trans. Med. Imaging*, vol. 30, no. 5, pp. 1028-1041, 2010.

[22] J. Caballero, A. N. Price, D. Rueckert, and J. V. Hajnal, "Dictionary learning and time sparsity for dynamic MR data reconstruction," *IEEE Trans. Med. Imaging*, vol. 33, no. 4, pp. 979-994, 2014.

[23] S. Ravishankar and Y. Bresler, "Sparsifying transform learning for compressed sensing MRI," in *Proc. IEEE 13th Int. Symp. Biomed. Imag.(ISBI)*, 2013, pp. 17-20.

[24] T. M. Quan, T. Nguyen-Duc, and W.-K. Jeong, "Compressed sensing MRI reconstruction using a generative adversarial network with a cyclic loss," *IEEE Trans. Med. Imaging*, vol. 37, no. 6, pp. 1488-1497, 2018.

[25] G. Yang *et al.*, "DAGAN: deep de-aliasing generative adversarial networks for fast compressed sensing MRI reconstruction," *IEEE Trans. Med. Imaging*, vol. 37, no. 6, pp. 1310-1321, 2017.

[26] Y. Han, J. Yoo, H. H. Kim, H. J. Shin, K. Sung, and J. C. Ye, "Deep learning with domain adaptation for accelerated projection-reconstruction MR," *Magn. Reson. Med.*, vol. 80, no. 3, pp. 1189-1205, 2018.

[27] D. Lee, J. Yoo, S. Tak, and J. C. Ye, "Deep residual learning for accelerated MRI using magnitude and phase networks," *IEEE Trans. Biomed. Eng.*, vol. 65, no. 9, pp. 1985-1995, 2018.

[28] M. Mardani *et al.*, "Deep generative adversarial neural networks for compressive sensing MRI," *IEEE Trans. Med. Imaging*, vol. 38, no. 1, pp. 167-179, 2018.

[29] Y. Han, L. Sunwoo, and J. C. Ye, "k-space deep learning for accelerated MRI," *IEEE Trans. Med. Imaging*, 2019.

[30] Y. Yang, J. Sun, H. Li, and Z. Xu, "ADMM-CSNet: A Deep Learning Approach for Image Compressive Sensing," *IEEE Trans. Pattern Anal. Mach. Intell.*, 2018.

[31] C. Qin, J. Schlemper, J. Caballero, A. N. Price, J. V. Hajnal, and D. Rueckert, "Convolutional recurrent neural networks for dynamic MR image reconstruction," *IEEE Trans. Med. Imaging*, vol. 38, no. 1, pp. 280-290, 2018.

[32] H. K. Aggarwal, M. P. Mani, and M. Jacob, "Modl: Model-based deep learning architecture for inverse problems," *IEEE Trans. Med. Imaging*, vol. 38, no. 2, pp. 394-405, 2018.

[33] J. Schlemper *et al.*, "Stochastic deep compressive sensing for the reconstruction of diffusion tensor cardiac MRI," in *International conference on medical image computing and computer-assisted intervention*, 2018: Springer, pp. 295-303.

[34] J. Schlemper, J. Caballero, J. V. Hajnal, A. N. Price, and D. Rueckert, "A deep cascade of convolutional neural networks for dynamic MR image reconstruction," *IEEE Trans. Med. Imaging,* vol. 37, no. 2, pp. 491-503, 2017.

[35] B. Zhu, J. Z. Liu, S. F. Cauley, B. R. Rosen, and M. S. Rosen, "Image reconstruction by domain-transform manifold learning," *Nature,* vol. 555, no. 7697, p. 487, 2018.

[36] S. Wang *et al.*, "DIMENSION: Dynamic MR imaging with both k-space and spatial prior knowledge obtained via multi-supervised network training," *NMR Biomed.*, 2018.

[37] R. Souza, R. M. Lebel, and R. Frayne, "A Hybrid, Dual Domain, Cascade of Convolutional Neural Networks for Magnetic Resonance Image Reconstruction," *Proceedings of Machine Learning* vol. 1, p. 11, 2019.

[38] T. Eo, Y. Jun, T. Kim, J. Jang, H. J. Lee, and D. Hwang, "KIKI‐net: cross‐domain convolutional neural networks for reconstructing undersampled magnetic resonance images," *Magn. Reson. Med.*, vol. 80, no. 5, pp. 2188-2201, 2018.

[39] K. Hammernik *et al.*, "Learning a variational network for reconstruction of accelerated MRI data," *Magn. Reson. Med.*, vol. 79, no. 6, pp. 3055-3071, 2018.

[40] S. Wang *et al.*, "Accelerating magnetic resonance imaging via deep learning," in *Proc. IEEE 13th Int. Symp. Biomed. Imag.(ISBI)*, 2016, pp. 514-517.

[41] K. He, X. Zhang, S. Ren, and J. Sun, "Deep residual learning for image recognition," in *Proceedings of the IEEE conference on computer vision and pattern recognition(CVPR)*, 2016, pp. 770-778.

[42] D. Kinga and J. B. Adam, "Adam: A method for stochastic optimization," in *Proc. Int. Conf. Learning Representations (ICLR)*, 2015, vol. 5.

[43] X. Qu, Y. Hou, F. Lam, D. Guo, J. Zhong, and Z. Chen, "Magnetic resonance image reconstruction from undersampled measurements using a patch-based nonlocal operator," *Med. Image Anal.*, vol. 18, no. 6, pp. 843-856, 2014.

[44] M. Seitzer *et al.*, "Adversarial and perceptual refinement for compressed sensing MRI reconstruction," in *Int. Conf. Med. Image Comput. Comput. Assist. Interv.(MICCAI)*, 2018: Springer, pp. 232-240.

[45] H. Chen *et al.*, "LEARN: Learned experts' assessment-based reconstruction network for sparse-data CT," *IEEE Trans. Med. Imaging,* vol. 37, no. 6, pp. 1333-1347, 2018.

[46] C. Trabelsi *et al.*, "Deep complex networks," *arXiv preprint arXiv:1705.09792*, 2017.